\documentclass{appolb}
\usepackage{graphicx}
\usepackage{hyperref}
\usepackage{cite}
\begin{document}
\title{Center Vortices and Topological Charge\thanks{Invited Talk at Excited QCD 2017 in Sintra, Portugal}}
\author{Roman H\"ollwieser}
\maketitle
\vspace{-8mm}
\begin{abstract}
I review important aspects of the relation between center vortices and topological charge, leading to chiral symmetry breaking.
\end{abstract}
  
Center vortices~\cite{'tHooft:1977hy,Vinciarelli:1978kp,Yoneya:1978dt,Cornwall:1979hz,Mack:1978rq,Nielsen:1979xu} are promising candidates for explaining confinement. They form closed magnetic flux tubes, whose flux is quantized, taking only values in the center of the gauge group. These properties are the key ingredients in the vortex model of confinement, which is theoretically appealing and was also confirmed by a multitude of numerical calculations, both in lattice Yang-Mills theory and within a corresponding infrared effective model, see {\textit e.g.}, \cite{DelDebbio:1996mh,Langfeld:1997jx,DelDebbio:1997ke,Langfeld:1998cz,Kovacs:1998xm,Engelhardt:1999wr,Engelhardt:1999fd,Bertle:2002mm,Engelhardt:2003wm,Hollwieser:2014lxa,Altarawneh:2015bya,Hollwieser:2015qea,Altarawneh:2016ped}, or~\cite{Greensite:2003bk}, which summarizes the main features. 

Vortices that randomly penetrate a given Wilson loop very naturally give rise to an area law. Since vortices are closed surfaces, the necessary randomness can be facilitated only by large vortices. This is further translated into the percolation of vortices, meaning that the size of the (largest) vortex clusters becomes comparable to the extension of the space itself. This percolation has been observed in lattice simulations of the confined phase, see {\textit e.g.}, \cite{Langfeld:1997jx}, while in the deconfined phase the vortices align in the time-like direction and the percolation mechanism remains working only for spatial Wilson loops \cite{Langfeld:1998cz,Engelhardt:1999fd}. This parallels percolation properties of monopoles. Moreover, it conforms with the observation at high temperatures that the spatial Wilson loops keep a string tension in contrast to the correlators of Polyakov loops.

Due to the color screening by gluons the string tension of pairs of static color charges in $SU(N)$ gauge theories depends on their $N$-ality. From the field perspective this $N$-ality dependence has its origin in the gauge field configurations which dominate the path integrals in the infrared. Center vortices are the only known configurations with appropriate properties. Recent results~\cite{Greensite:2014gra} have also suggested that the center vortex model of confinement is more consistent with lattice results than other currently available models. If one considers that a phase transition of the gauge field influences both gluons and fermions, then one would expect that deconfinement and chiral phase transition are directly related and rely on the same mechanism. 

Lattice studies indicate that vortices are also responsible for topological charge~\cite{Engelhardt:2000wc,Bertle:2001xd,Bruckmann:2003yd,Jordan:2007ff,Engelhardt:2010ft,Hollwieser:2010mj,Hollwieser:2011uj,Schweigler:2012ae,Hollwieser:2012kb,Hollwieser:2014mxa,Nejad:2015aia,Hollwieser:2015koa} and chiral symmetry breaking~\cite{deForcrand:1999ms,Alexandrou:1999vx,Engelhardt:1999xw,Reinhardt:2000ck,Engelhardt:2002qs,Leinweber:2006zq,Bornyakov:2007fz,Hollwieser:2008tq,Bowman:2010zr,Hollwieser:2013xja,Hoellwieser:2014isa,Trewartha:2014ona,Trewartha:2015nna}, and thus unify all non-perturbative phenomena engendered by the structure of the strong interaction vacuum in a common framework. 

In~\cite{Engelhardt:1999xw,Reinhardt:2001kf} it was shown that center vortices, quantized magnetic fluxes in the QCD vacuum, contribute to the topological charge by intersections with $Q=\pm1/2$ and writhing points with a value of $\pm1/16$.  By the Atiyah-Singer index theorem~\cite{Atiyah:1971rm,Schwarz:1977az,Brown:1977bj,Narayanan:1994gw} zero modes are related to one unit of topological charge. Therefore, the question emerges, how vortex intersections and writhing points are related to these zero modes. Ref.~\cite{Hollwieser:2011uj} compares vortex intersections with the distribution of zero modes of the Dirac operator in the fundamental and adjoint representation using both the overlap and AsqTad staggered fermion formulations in SU(2) lattice gauge theory. By forming arbitrary linear combinations of zero modes they prove that their scalar density peaks at least at two intersection points~\cite{Hollwieser:2011uj}. Further, since it is expected that zero modes of the Dirac operator concentrate in regions of large topological charge density, a correlation between the location of vortex intersections and writhing points and the density $\rho_\lambda(x)=|\psi_\lambda(x)|^2$ of eigenmodes of the Dirac operator $D$, where $D\psi_\lambda=\lambda\psi_\lambda$ with $\lambda=0$ in the overlap formulation and $\lambda\approx0$ in the AsqTad staggered formulation supports this picture~\cite{Hollwieser:2008tq,Kovalenko:2005rz}. The correlation is strong for zero- and low-lying modes, and decreases for higher eigenmodes. The authors of \cite{Kovalenko:2005rz} further conclude that the eigenmode correlation on two-dimensional surfaces (vortices) is stronger than for three-dimensional objects. A positive (low-lying) Dirac eigenmode - vortex structure resp. topological charge correlation is a first indication of the importance of center vortices for chiral symmetry breaking.

In recent investigations further sources of topological charge from center vortices were discovered. Colorful spherical SU(2) vortices \cite{Jordan:2007ff,Hollwieser:2010mj,Schweigler:2012ae,Hollwieser:2012kb} and colorful plain vortices \cite{Nejad:2015aia,Nejad:2016fcl} were introduced. They contributes to the topological charge by their color structure and attract zeros mode like instantons. A spherical vortex can be constructed in one time-slice $t_0$, by putting a hedgehog-like gauge field on {\textit e.g.} the time-like links $U_4(\vec r,t_0)=\mathrm e^{\mathrm i\alpha(r)\vec n(\vec r)\vec\sigma}\in SU(2)$ around a two-dimensional sphere in $R^3$. $\alpha(r)$ varies monotoniously between $0$ and $\pi$ from the center $r=0$ of the sphere to large distances and the color direction is chosen as $\vec n=\vec r/r$. As the field at large distances is independent of the direction this ball in $R^3$ is isomorphic to $S^3\simeq SU(2)$ and characterized by a winding number $N=\pm1$ for the spherical vortices and accordingly in a topological charge via the Atiyah-Singer index theorem. In \cite{Schweigler:2012ae}, the continuum object corresponding to the spherical vortex was identified by a gauge transformation transferring the topological structure from the time-like links to the corresponding space-like links. After this gauge transformation the spherical vortex can be distributed over several time slices and was identified as vacuum to vacuum transition in temporal direction and its similarity to an instanton was demonstrated. Similarly color structures were introduced in \cite{Nejad:2015aia,Nejad:2016fcl} on plain vortex structures.

Further, in \cite{Hollwieser:2012kb}, it was shown how the interplay of various topological structures from center vortices (and instantons) leads to near-zero modes, which by the Banks-Casher relation~\cite{Banks:1979yr} are responsible for a finite chiral condensate, using the overlap and AsqTad staggered Dirac operators. The spectra of (anti-/)instantons, spherical (anti-/)vortices and pairs of two of these individual objects on otherwise trivial lattice configurations are show nearly exactly the same eigenvalues for instantons and spherical vortices, as well as different pairs, single objects attract a zero mode whereas would-be zero modes of two objects in one lattice configuration result in a pair of near-zero modes, consisting of two chiral parts corresponding to the two constituents of the pairs. Similar results were presented for vortex intersections from plane vortices, which due to their topological charge contributions give rise of zero and near-zero modes~\cite{Hollwieser:2011uj}.

These observations lead to a picture similar to the instanton liquid model. The lumps of topological charge appearing in Monte-Carlo configurations interact in the QCD-vacuum and determine the density of near-zero modes. Therefore, it is not the true zero modes deciding on the value of the topological charge of a field configuration which lead to the breaking of chiral symmetry. The number of these modes is small in the continuum limit. It is the density of interacting topological objects which leads to the density of modes around zero and according to the Banks-Casher relation determines the strength of chiral symmetry breaking.

In the vortex picture the model of chiral symmetry breaking can be formulated even more generally, as it was shown that various shapes of vortices attract (would-be) zero modes which contribute via interactions to a finite density of near-zero modes with local chiral properties, {\textit i.e.}, local chirality peaks at corresponding topological charge contributions. In Monte-Carlo configurations, there are no perfectly flat or spherical vortices, as one does not find perfect instantons, but the general picture of topological charge from vortex intersections, writhing points and even color structure contributions or instantons can provide a general picture of $\chi$SB: any source of topological charge can attract (would-be) zero modes and produce a finite density of near-zero modes leading to chiral symmetry breaking via the Banks-Casher relation. Here one also has to ask what could be the dynamical explanation of $\chi$SB. One can try the conjecture that only a combination of color electric and magnetic fields leads to $\chi$SB, electric fields accelerating color charges and magnetic fields trying permanently to reverse the momentum directions on spiral shaped paths. Such reversals of momentum keeping the spin of the particles should especially happen for very slowly moving color charges. Alternatively one could argue that magnetic color charges are able to flip the spin of slow quarks, {\textit i.e.}, when they interact long enough with the vortex structures.


The Adelaide group~\cite{OMalley:2011aa,Bowman:2010zr,Trewartha:2014ona,Trewartha:2015nna,Bowman:2008qd,O'Malley:2011zz,Trewartha:2015ida,Kamleh:2017lij} shows that center vortices underpin both, confinement and dynamical chiral symmetry breaking in $SU(3)$ lattice gauge theory. They look at the topological charge density, the static quark potential, the quark mass function, and the hadron mass spectrum on original (untouched), vortex-only and vortex-removed ensembles. 
The background of instanton-like objects emerging from the vortex-only configurations under cooling is examined in \cite{Trewartha:2015ida} by examining the local maxima of the action density. It is shown that after just 10 sweeps of smoothing the local maxima stabilize and begin to resemble classical instantons in shape and corresponding topological charge density at the center \cite{Moran:2008qd}. 
The number of instanton-like objects found on original and vortex-only configurations remains about equal even after large amounts of cooling, whereas the number of objects on vortex-removed configurations is greatly reduced. Thus, while vortex-removal destabilizes the otherwise topologically non-trivial instanton-like objects it is possible to create an instanton liquid-like background on vortex-only configurations, analogous to that found on Monte-Carlo generated configurations after similar smoothing. Through calculations of the static quark potential and Landau-gauge overlap propagator, it was shown that this background is able to reproduce all salient long-range features of the original configurations. Therefore, the information necessary to recreate the long-range structure of the QCD vacuum is contained within the center vortex degrees of freedom.

The importance of the long-range nature of low-dimensional topological structures for the understanding of the mechanism of $\chi$SB in QCD was also underlined by Buividovich {\textit et. al}~\cite{Horvath:2002zy,Horvath:2003is,Braguta:2010ej,Buividovich:2011cv,Braguta:2013kpa} and agrees well with a vortex picture of $\chi$SB. Since the QCD-vacuum is strongly non-perturbative, it does not contain semiclassical instantons~\cite{Horvath:2002gk,Horvath:2004gw} but is crowded with topologically charged objects which, after smooth reduction of the action (also known as cooling), may become instantons. In pure $SU(3)$ lattice gauge theory in a typical equilibrium configuration about 80\% of space-time points are covered by two oppositely-charged connected structures built of elementary three-dimensional coherent hypercubes. The hypercubes within the structure are connected through two-dimensional common faces suggesting that this coherence is a manifestation of a low-dimensional order present in the QCD vacuum~\cite{Horvath:2002zy,Horvath:2003is}. Ref.~\cite{Buividovich:2011cv} analyzes the localization properties of fermionic zero modes and demonstrates that topological charge and chirality are localized on structures with fractal dimension $2\le D\le3$, favoring the vortex/domain-wall nature of the localization~\cite{Engelhardt:1999xw,Polikarpov:2004iv,Kovalenko:2004xm}. 

\bibliographystyle{utphys}
\bibliography{chiral}

\end{document}